\documentclass[]{interact}

\usepackage{epstopdf}
\usepackage[caption=false]{subfig}

\usepackage[numbers,sort&compress,merge]{natbib}
\bibpunct[, ]{[}{]}{,}{n}{,}{,}
\renewcommand\bibfont{\fontsize{10}{12}\selectfont}

\theoremstyle{plain}

\theoremstyle{definition}

\theoremstyle{remark}

\usepackage{IEEEtrantools}
\newcommand{\ai}{\emph{ab initio}}
\newcommand{\EinA}{Einstein-\emph{A} coefficients}

\begin{document}


\title{Manifestation of the Normal Intensity Distribution Law (NIDL) in the rovibrational emission spectrum of hydroxyl radical}

\author{
\name{Emile~S.~Medvedev\textsuperscript{a}\thanks{CONTACT: E.~S.~Medvedev. Email: medvedev@icp.ac.ru.}, Aleksander~Yu.~Ermilov\textsuperscript{b}, and Vladimir~G.~Ushakov\textsuperscript{a}}
\affil{\textsuperscript{a}Federal Research Center of Problems of Chemical Physics and Medicinal Chemistry (former Institute of Problems of Chemical Physics), Russian Academy of Sciences, 142432 Chernogolovka, 1 Prospekt Akademika N.~N.~Semenova, Russian Federation; \textsuperscript{b}M.~V.~Lomonosov Moscow State University, 119991 Moscow, Leninskie Gory 1/3, Russian Federation}
}

\maketitle

\begin{abstract}
    The latest experimental [Noll \emph{et al.} Atmos. Chem. Phys. 20(2020)5269] and theoretical [Brooke \emph{et al.} J. Quant. Spectr. Rad. Transfer 168(2016)142] data on the OH emission intensities are analyzed with use of the NIDL. It is found that the calculated intensities of the $\Delta v>6$ transitions should not be trusted.  
    The analysis of the OH data revealed that the NIDL theory is not applicable to the satellite bands. The effect of small reduced mass previously discovered in H$_2$ [Ushakov \emph{et al.} J. Mol. Spectrosc. 399(2024)111863], causing the NIDL straight-line slope to be larger than the one associated with the repulsive branch of the potential, is demonstrated in OH, and the same should be true of all the diatomic hydrides. We performed \ai\ calculations of the OH repulsive branch and compared it with the one of Brooke \emph{et al.} and the other due to Varandas and Voronin [Chem. Phys. 194(1995)91]. We found that the \ai\ PEF closely follows the Varandas-Voronin potential in the repulsive region important for calculating the overtone intensities [Medvedev, J. Chem. Phys. 137(2012)174307]. Assumption is made  that different potentials should be used to calculate the transition frequencies and the intensities of the overtone bands for spectroscopic databases.
\end{abstract}


\section{Introduction}

More than thirty years ago, two of us in collaboration with our beautiful friend and colleague Aleksander Nemukhin published paper \cite{Ermilov90} in support of the NIDL theory (see review \cite{Medvedev12} and references therein). The theory has been verified by the available experimental and theoretical data for a number of diatomic molecules and quasi-diatomic local vibrations in polyatomic molecules, and it has proven to be a powerful tool to control the precision of the calculated intensities of the overtone transitions \cite{Li15,Ushakov23PN}. In particular, the OH observational data of Krassovsky \emph{et al.} \cite{Krassovsky61} and Cosby and Slanger \cite{Cosby07} were used to demonstrate the NIDL behavior of the relative emission intensities \cite{Medvedev85,Medvedev12}.

In this paper, we use the contemporary data on hydroxyl radical \cite{Brooke16,Noll20,Gordon22} to verify the NIDL for emission and to demonstrate the utility of the NIDL for the analysis of the calculated overtone intensities.

We perform the \ai\ calculations of the PEF repulsive branch important for calculations of the overtone intensities \cite{Medvedev12}. The calculated repulsive branch is compared with the ones of two literature PEFs to determine which one is more suited for calculations of the overtone intensities.

Section \ref{pre} gives a brief review of the predictions of the NIDL theory for emission. Sections \ref{ver13} and \ref{ver4} provide for verifications of these predictions in the emission spectra. Section \ref{ai} describes the \ai\ calculations and their significance for the calculations of the overtone intensities. Section \ref{conclusions} summarize our findings.

\section{Predictions of the NIDL theory}\label{pre}

The NIDL theory is based on the quasi-classical approximation, which states that, at high-enough energies, the vibrational  wave function can be represented in the form $p^{-1/2}\exp{\left(iS/\hbar \right)}$, where $p$ and $S$ are classical momentum and action in a given vibrational state, $E_v$, which is formally equivalent to $v\gg1$. In practice, however, $v\ge2$ is sufficient.\footnote{Actually, even $v=0$ and 1 can be treated quasi-classically because the NIDL formalism considers the wave functions in the complex plane far enough from the turning points where the quasi-classical behavior is assured, but we leave aside this issue since it is beyond the scope of the present paper.}

Another important feature of the theory is application of the Franck-Condon principle, which states that, due to a large difference between the electron and nuclear masses, any optical transition at frequency $\nu$ between states 1 and 2 occurs at a fixed nuclear configuration, $r^\star$, where the nuclear momenta coincide, $p_1(r^\star)=p_2(r^\star)$ and the potentials differ by the photon energy, $U_1(r^\star)=U_2(r^\star)+h\nu$. 

In application to the rovibrational transitions within the ground electronic state, where $U_1(r)=U_2(r)\equiv U(r)$, the Franck-Condon principle means \cite[Secs. 5.4-5.6]{Medvedev95} that the main contribution to the transition-dipole-moment (TDM) integral is provided by a vicinity of point $r^\star$ in the complex plane  where the potential-energy function (PEF) has singularity, $U(r)\rightarrow\infty$, see \S51 in Ref. \cite{Landau77}. 

There is one and only one \emph{physical} singularity of $U(r)$, namely that at $r=0$, due to the Coulomb nuclear repulsion,\footnote{This important notion means that any model theoretical PEF must not have any other singularities affecting the TDMs.} therefore the repulsive branch of $U(r)$ plays a crucial role in determining the overtone intensities. If we approximate the repulsive branch with a simple exponential function, $U(r)\propto\exp{\left(-2\beta r \right)}$, then the TDM squared\footnote{For brevity, we will call the product, $A\lambda^3\propto\textrm{TDM}^2$, the intensity.} for the overtone transition $\left(\Delta v\equiv v^\prime-v^{\prime\prime}\ge2 \right)$ from the upper level $v^\prime$ to the lower level $v^{\prime\prime}$ obeys the NIDL,
\begin{equation}
    \log \textrm{TDM}^2_{v^\prime v^{\prime\prime}} = \textrm{const} - a\sqrt{E_{v^{\prime}}/\omega}, \label{NIDLa}
\end{equation}
where the energy and harmonic frequency of vibration, $\omega$, are in cm$^{-1}$, and the upper level is assumed to be high, $v^\prime\ge2$, which occurs in both absorption and emission. The const is assumed to be a slow function of $v^\prime$ at a given $v^{\prime\prime}$ \footnote{But see below, Sec. \ref{ver4}.} except for the anomalies \cite[Sec. II]{Medvedev12}, which do not obey the NIDL and must be excluded from the data fitting.

If the lower level is also high, $v^{\prime\prime}\ge2$, which is often met in emission, then the NIDL for the ratio of the intensities, TDM$^2$, for two overtone transitions starting at a common upper level takes the form (we omit primes for brevity)
\begin{equation}
    \log \left( \frac{\textrm{TDM}_{vv_1}}{\textrm{TDM}_{vv_2}}\right)^2  = a\left(\sqrt{\frac{E_{v_1}}{\omega}}-\sqrt{\frac{E_{v_2}}{\omega}}\right), \hspace{10pt} v_1> v_2\ge2, \hspace{10pt}v-v_1\ge2, \label{NIDLe}
\end{equation}
where $a$ is the same as in Eq. (\ref{NIDLa}). Here, the const disappears because the left-hand side vanishes at $v_1=v_2$. In fact, due to approximate nature of the NIDL, there is a small const, on the order of the statistical error, which, however, becomes large at the anomalies; yet, the anomalies are ignored in the NIDL plots.

Finally, if the lowest level is 0 or 1, equation (\ref{NIDLe}) is modified as
\begin{equation}
    \log \left( \frac{\textrm{TDM}_{vv_1}}{\textrm{TDM}_{vv_2}}\right)^2  = \textrm{const} + a\sqrt{\frac{E_{v_1}}{\omega}}, \hspace{10pt}  v_2=0,1 \label{NIDLe10}
\end{equation}
(for more details, see review \cite{Medvedev12} and references therein). 

Graphically, Eqs. (\ref{NIDLa})-(\ref{NIDLe10}) are represented, in the respective coordinates, by straight lines with slope $a$, which is connected to the steepness, $\beta$, of the repulsive branch of the PEF by relation \cite{Medvedev12}
\begin{equation}
    a = \frac{\pi}{\beta r_\textrm{e}\ln{10}}\sqrt{\frac{\omega}{B_\textrm{e}}}, \label{a}
\end{equation}
where $r_\textrm{e}$ is equilibrium bond length, harmonic frequency $\omega$ and rotational constant $B_\textrm{e}$ are in cm$^{-1}$.

There are several consequences of the above equations that are easy to verify. Equation (\ref{NIDLa}) predicts the exponential decrease of the intensities with the overtone number, of which the pace, $a$, given by Eq. (\ref{a}) is inversely proportional to the steepness of the repulsive branch of the PEF: the steeper the PEF, the slowlier the decay. The decay rate in Eq. (\ref{a}) depends solely on the PEF, being independent of the dipole-moment function (DMF), which affects only the const.\footnote{See, however, Sec. \ref{ver4}; the DMF is also responsible for the anomalies, which drop off the NIDL line.} 

Equation (\ref{NIDLe}) predicts that the ratio of the intensities of two lines emitted from a common upper level, $v$, is the same for various $v$. This important feature of the overtone transitions was first noted by Ferguson and Parkinson \cite{Ferguson63}, who even considered a possibility to use the linear DMF (``the relative
intensities in the high overtone sequences are likely to be similar to those for a linear dipole
moment").

Finally, Eq. (\ref{a}) permits direct verification of the NIDL theory by calculating $\beta$ with the \ai\ methods and comparing it with the one derived from the NIDL slope. 

Here, we will use the theoretical \cite{Brooke16} and experimental \cite{Noll20} data to verify the NIDL and to derive the $\beta$ values from the NIDL slopes; then we perform the \ai\ calculations to compare the above $\beta$ with the steepness of the repulsive branches of both the present \ai\ PEF and the literature PEFs.

\section{Verification of Eqs. (\ref{NIDLa})-(\ref{NIDLe10})}\label{ver13}

\begin{figure}[htbp]
    \centering
    \includegraphics[scale=0.4]{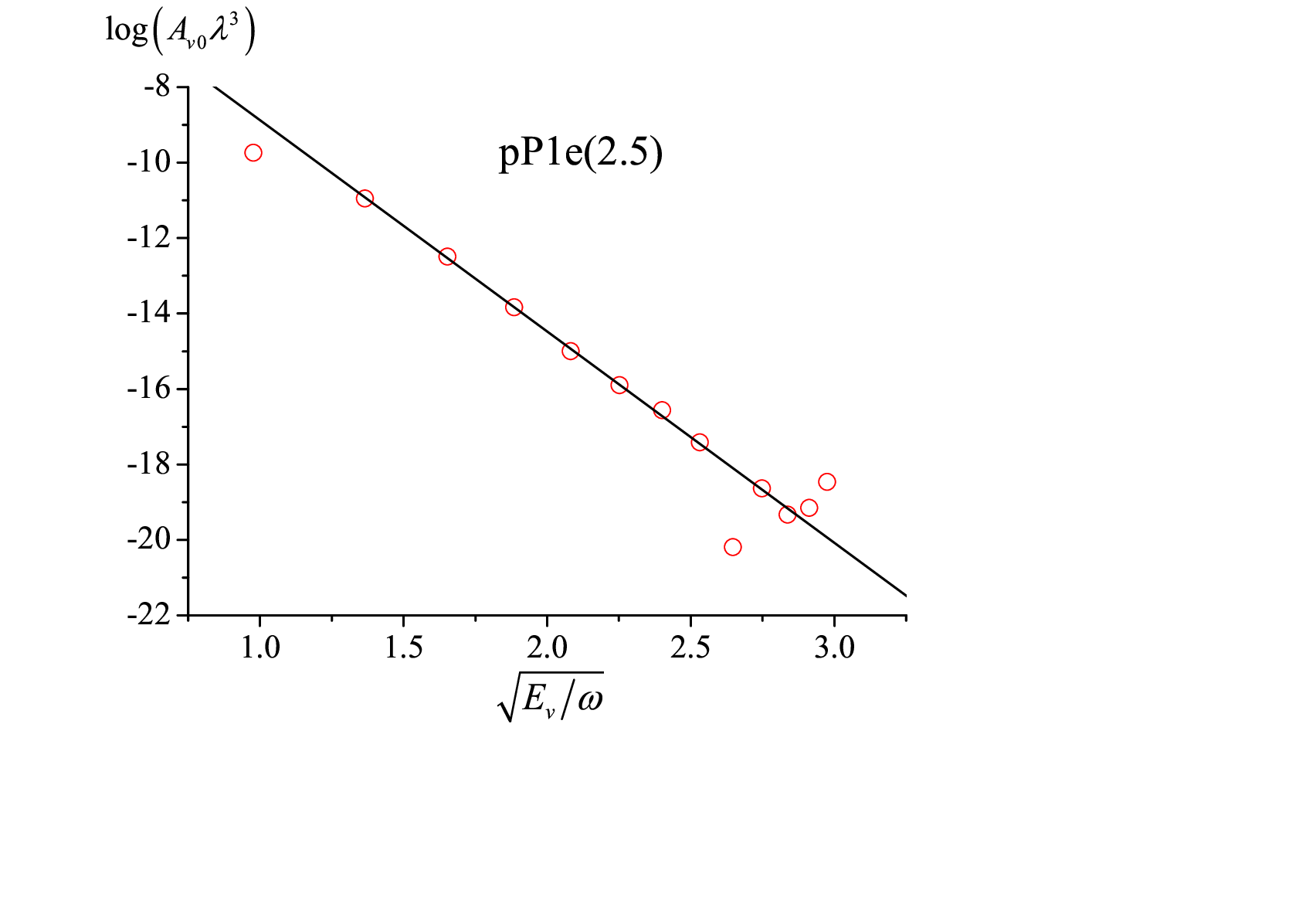}
    \includegraphics[scale=0.4]{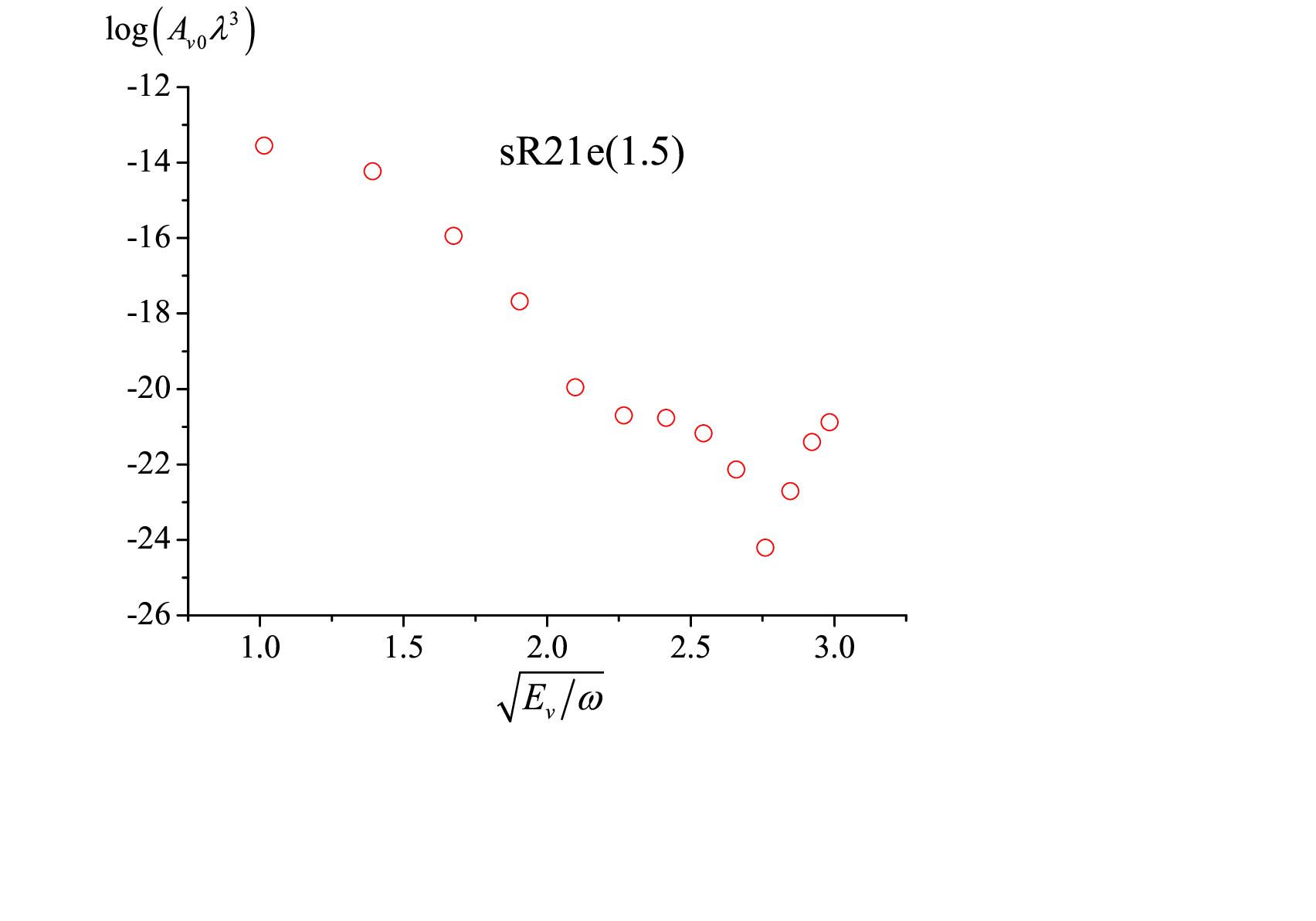}
    \caption{\EinA\ for the main-band low-$J$ $v$-0 transitions, $v=1$-13, from Ref. \cite{Brooke16} normalized to the frequency factor to obtain TDM$^2$. The NIDL line in the upper panel is plotted by least-squares fitting of Eq. (\ref{NIDLa}) to the $v=2$-8 intensities. Data at $v>8$ are unreliable. The lower panel shows a wavy line, which testifies that the NIDL fails for the intercombination lines.}
    \label{pP1e25eq1}
\end{figure}

When plotting the data in the NIDL coordinates according to Eq. (\ref{NIDLa}), we discovered that only the main-band (intra-multiplet, $\Delta F=0$) transitions obey the NIDL. This is illustrated in Fig. \ref{pP1e25eq1}, where the upper and lower panels show examples of the intra- and intercombination ($\Delta F\ne0$, satellite) lines. The reason for this different behavior is explained by 
the fact that the very existence of the satellite transitions is due to 
the rotational mixing of two $X$ multiplet sub-states with different potentials. 
Moreover, severe cancellation of various contributions to the TDM for the satellite transitions takes place \cite{Mies74,Nelson89a,Ushakov24a}, hence this is a special case not covered by the NIDL theory.
We remind that the anomalies also do not obey the NIDL, hence we have here a second reason for the NIDL to fail.

Table \ref{tab:eq1} shows the NIDL slopes for a few main-band low-$J$ transitions.\footnote{The NIDL theory was developed for purely vibrational transitions, therefore only low $J$ are in order to consider to minimize the effect of rotation.} 
Previously \cite{Medvedev85}, it was found that $\beta=3.57\pm0.22$ \AA$^{-1}${} from the NIDL slope, $a=5.54\pm0.33$, derived from the OH data by Krassovsky \emph{et al.} \cite{Krassovsky61}; a similar result was obtained in Ref. \cite{Medvedev12} from the HITRAN 2008 data. It is seen from the table that the NIDL slopes are very close to the one cited above. 

\begin{table}[htbp]
    \centering
    \caption{The NIDL slopes derived from fitting the data of \textbf{Brooke16} \cite{Brooke16} to Eq. (\ref{NIDLa}). The NIDL line is drawn across the $v=2$-8 points.}
    \vspace{5pt}
    \begin{tabular}{c|c}
    \hline
    Line & $a$  \\
    \hline
     pP1e(2.5)    & $5.60\pm0.10$ \\
     pP2e(2.5)    & $5.59\pm0.12$ \\
     qQ1e(1.5)	 & $5.61\pm0.13$ \\
     qQ2e(1.5)	 & $5.84\pm0.19$ \\	
     rR1e(2.5)    & $5.69\pm0.17$ \\
     rR2e(2.5)    & $5.74\pm0.17$ \\
     \hline
     average  & $5.68\pm0.15$ \\
     \hline
     Ref. \cite{Medvedev12} & $5.22\pm0.03^{a,b}$ \\
                            & $5.77\pm0.02^{a,c}$ \\
    \hline
    \multicolumn{2}{l}{$^a$Based on the absorption data from HITRAN 2008.} \\
    \multicolumn{2}{l}{$^b$From the plot of oscillator strength.} \\
    \multicolumn{2}{l}{$^c$From the plot of oscillator strength divided by frequency.} 
    \end{tabular}
    \label{tab:eq1}
\end{table}

In Fig. \ref{NIDLa}, the $v=9$ point looks like an anomaly, but in fact it manifests the beginning of chaotic behavior of the intensities, which is also confirmed for the other lines shown in Table \ref{tab:eq1}. Therefore, the intensities at $v>8$ are highly unreliable.

\begin{figure}[htbp]
    \centering
    \includegraphics[scale=0.4]{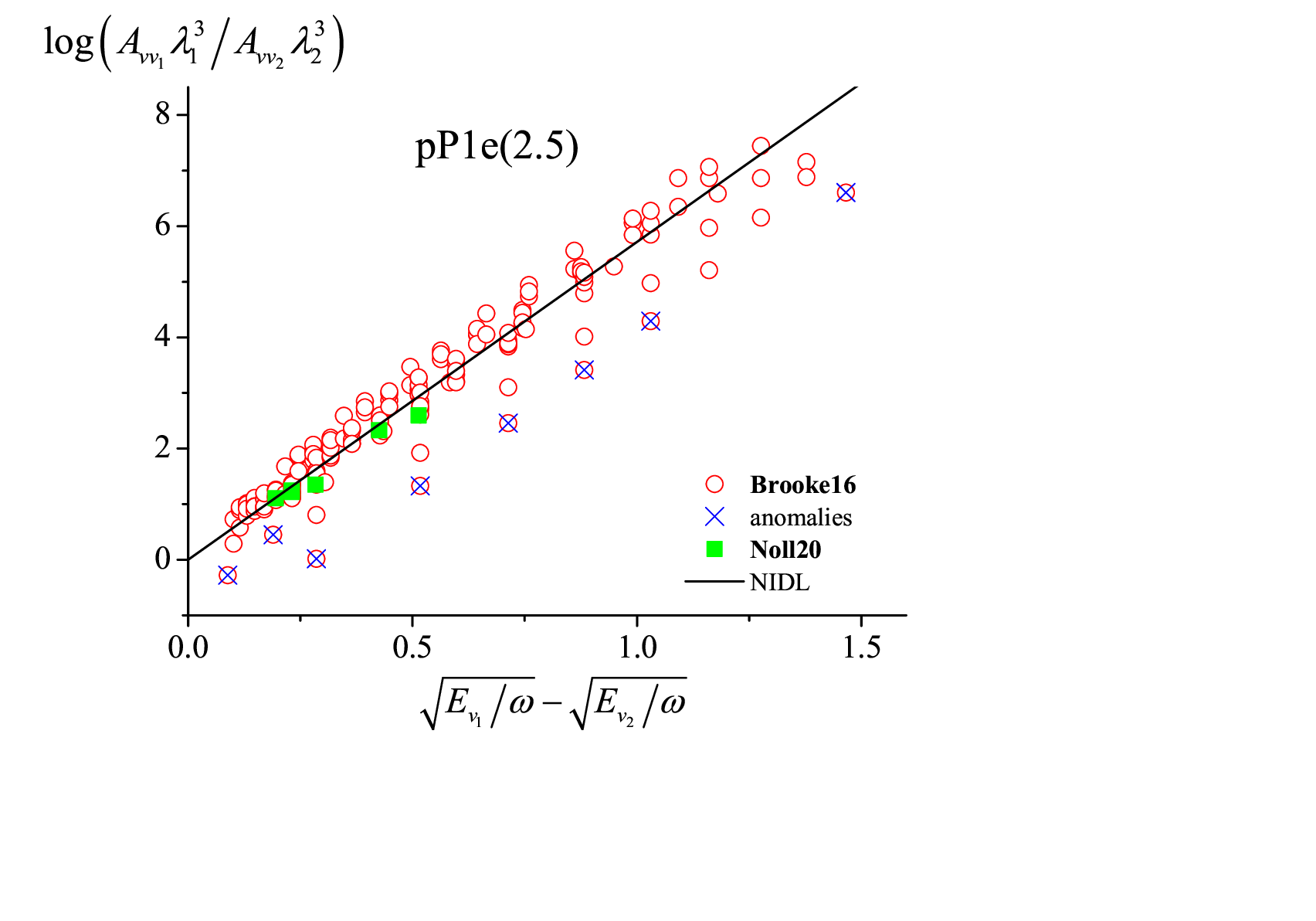}
    \includegraphics[scale=0.4]{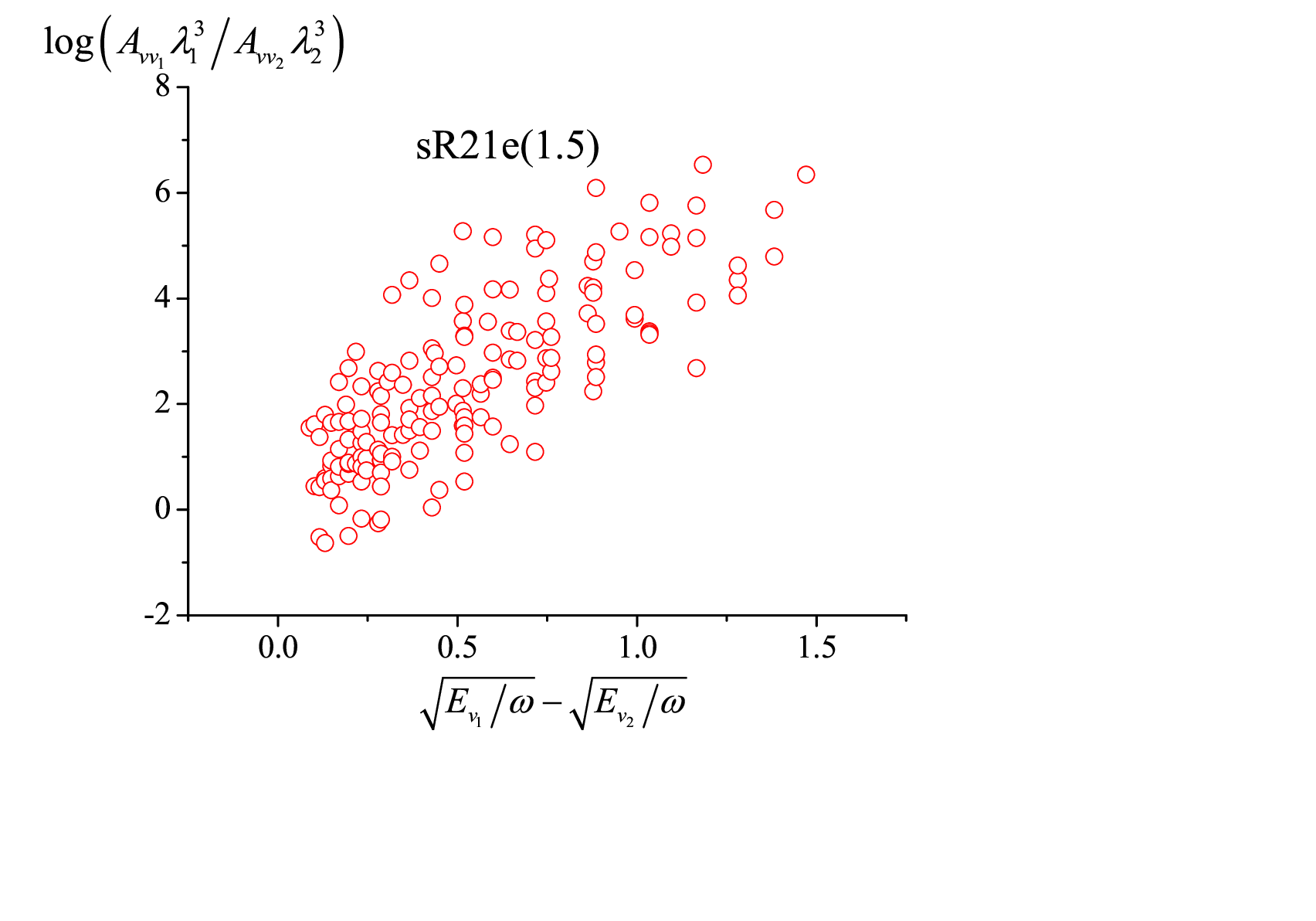}
    \caption{The relative intensities of the low-$J$ $v$-$v_1$ and $v$-$v_2$ transitions from a common upper level $v$ to lower levels $v_1>v_2\ge2$, $v\ge v_1+2$. Circles, data from \textbf{Brooke16} \cite{Brooke16}; crosses, the anomalies partly excluded$^\textrm{\ref{ft_an}}$ from the NIDL plot of Eq. (\ref{NIDLe}); squares, the observational data from \textbf{Noll20} \cite{Noll20}. The upper panel shows the main-band transitions, the lower the satellite transitions not observed in \textbf{Noll20}; for the latter, the NIDL fails because the variations along the ordinate at a given abscissa are huge.}
    \label{pP1e25eq2}
\end{figure}

Figure \ref{pP1e25eq2} shows the relative intensities of the lines emitted from a common upper level. In the upper panel (the main-band transitions), the NIDL is drawn by the least-squares fitting of Eq. (\ref{NIDLe}) to all data excluding some anomalies.\footnote{It is difficult to exclude all the anomalies. \label{ft_an}} The lower panel demonstrates the failure of the NIDL theory for the satellite transitions. Indeed, the intensity ratio must be relatively insensitive to the upper level, see Eq. (\ref{NIDLe}). In the figure, vertical sets of points are seen that correspond to transitions from various upper levels, $v$, to the same pair of the lower ones. Within each set, in contrast to the above prediction, the ratio varies dramatically, up to 5 orders of magnitude, which is comparable to the change of this ratio over the full range of abscissa (about 6 orders of magnitude); compare this with the upper panel where the variations within the vertical sets are much less, about 1.5 order of magnitude.

Table \ref{tab:eq2} presents the NIDL slopes of Eq. (\ref{NIDLe}) derived from the relative intensities of transitions with $v_2\ge2$ and $v$  up to the maximum value of $v_\textrm{max}$. The average value of the NIDL slope agrees with the one derived previously in Ref. \cite{Medvedev12} from the observational data of Krassovsky \emph{et al.} \cite{Krassovsky61} and of Cosby and Slanger \cite{Cosby07}.

\begin{figure}
    \centering
    \includegraphics[scale=0.4]{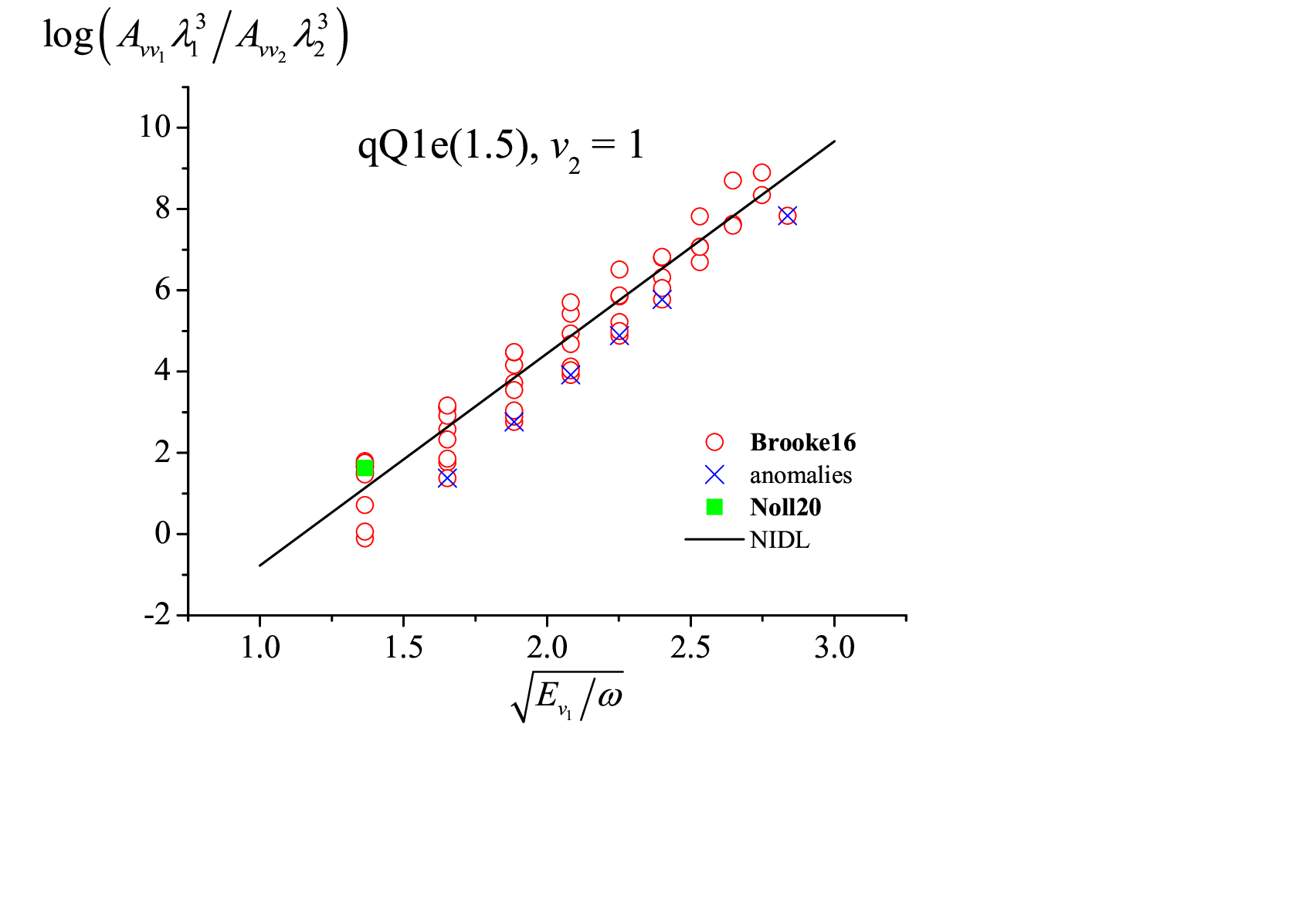}
    \caption{The same as in Fig. \ref{pP1e25eq2} with $v_2=1$.}
    \label{qQ1e15eq31}
\end{figure}

Figure \ref{qQ1e15eq31} shows an example of the NIDL plot according to Eq. (\ref{NIDLe10}) for transitions involving the lowest level $v_2=1$. The NIDL slopes derived for some other low-$J$ lines are collected in Table \ref{tab:eq3_1}. 

The data for transitions involving the lowest state $v_2=0$ are not shown as they contain too many anomalies that is not easy to exclude. The NIDL slope for this kind of transitions found in Ref. \cite{Medvedev85} is $5.62\pm0.48$.

\begin{table}[htbp]
    \centering
    \caption{The NIDL slopes derived from fitting the data of \textbf{Brooke16} \cite{Brooke16} to Eq. (\ref{NIDLe}).}
    \vspace{5pt}
    \begin{tabular}{c|c|c|c}
    \hline
    Line & $a$ & $v_\textrm{max}$ & No.$^\star$\\
    \hline
     pP1e(2.5)    & $5.72\pm0.29$ & 9 & 7 \\
     pP2e(2.5)    & $5.76\pm0.28$ & 9 & 5 \\
     qQ1e(1.5)	 & $5.73\pm0.29$ & 9 & 7 \\
     qQ2e(1.5)	 & $5.76\pm0.29$ & 9 & 7 \\	
     rR1e(1.5)    & $5.76\pm0.32$ & 9 & 0 \\
     rR2e(1.5)    & $5.82\pm0.33$ & 9 & 0 \\
     rR1e(2.5)    & $5.81\pm0.35$ & 9 & 1 \\
     rR2e(2.5)    & $5.88\pm0.38$ & 9 & 7 \\
    \hline
     \multicolumn{1}{c|}{average } &  \multicolumn{1}{c|}{$5.78\pm0.32$}\\
     \hline
     \multicolumn{1}{c|}{Ref. \cite{Medvedev85} } &  \multicolumn{1}{c|}{$5.54\pm0.33$}\\
     \multicolumn{1}{c|}{Ref. \cite{Medvedev12} } &  \multicolumn{1}{c|}{$5.36\pm0.22$}\\
    \hline
    \multicolumn{4}{l}{$^\star$The number of the \textbf{Noll20} \cite{Noll20}}\\
    \multicolumn{4}{l}{\hspace{5pt}experimental points in the plot.}
    \end{tabular}
    \label{tab:eq2}
\end{table}

\begin{table}[htbp]
    \centering
    \caption{The NIDL slopes derived from fitting the data of \textbf{Brooke16} \cite{Brooke16} to Eq. (\ref{NIDLe10}), $v_2=1$.}
    \vspace{5pt}
    \begin{tabular}{c|c|c|c}
    \hline
    Line & $a$ & $v_\textrm{max}$ & No.$^\star$\\
    \hline
     pP1e(2.5)    & $5.18\pm0.29$ & 10 & 1 \\
     pP2e(2.5)    & $5.22\pm0.28$ & 10 & 1 \\
     qQ1e(1.5)	 & $5.22\pm0.29$ & 10 & 1 \\
     qQ2e(1.5)	 & $5.17\pm0.29$ & 10 & 1 \\	
     rR1e(1.5)    & $5.23\pm0.32$ & 10 & 0 \\
     rR2e(1.5)    & $5.27\pm0.33$ & 10 & 0 \\
     rR1e(2.5)    & $5.25\pm0.35$ & 11 & 1 \\
     rR2e(2.5)    & $5.30\pm0.38$ & 10 & 1 \\
    \hline
     \multicolumn{1}{c|}{average } &  \multicolumn{1}{c|}{$5.22\pm0.32$}\\
     \hline
    \multicolumn{1}{c|}{Ref. \cite{Medvedev85} } &  \multicolumn{1}{c|}{$5.88\pm0.31$}\\
    \multicolumn{1}{c|}{Ref. \cite{Medvedev12} } &  \multicolumn{1}{c|}{$5.34\pm0.27$}\\
    \hline
    \multicolumn{4}{l}{$^\star$The number of the \textbf{Noll20} \cite{Noll20} }\\
    \multicolumn{4}{l}{\hspace{3pt} experimental points in the plot.}
    \end{tabular}
    \label{tab:eq3_1}
\end{table}

\section{Verification of Eq. (\ref{a})}\label{ver4}

\begin{figure}[htbp]
    \centering
    \includegraphics[scale=0.4]{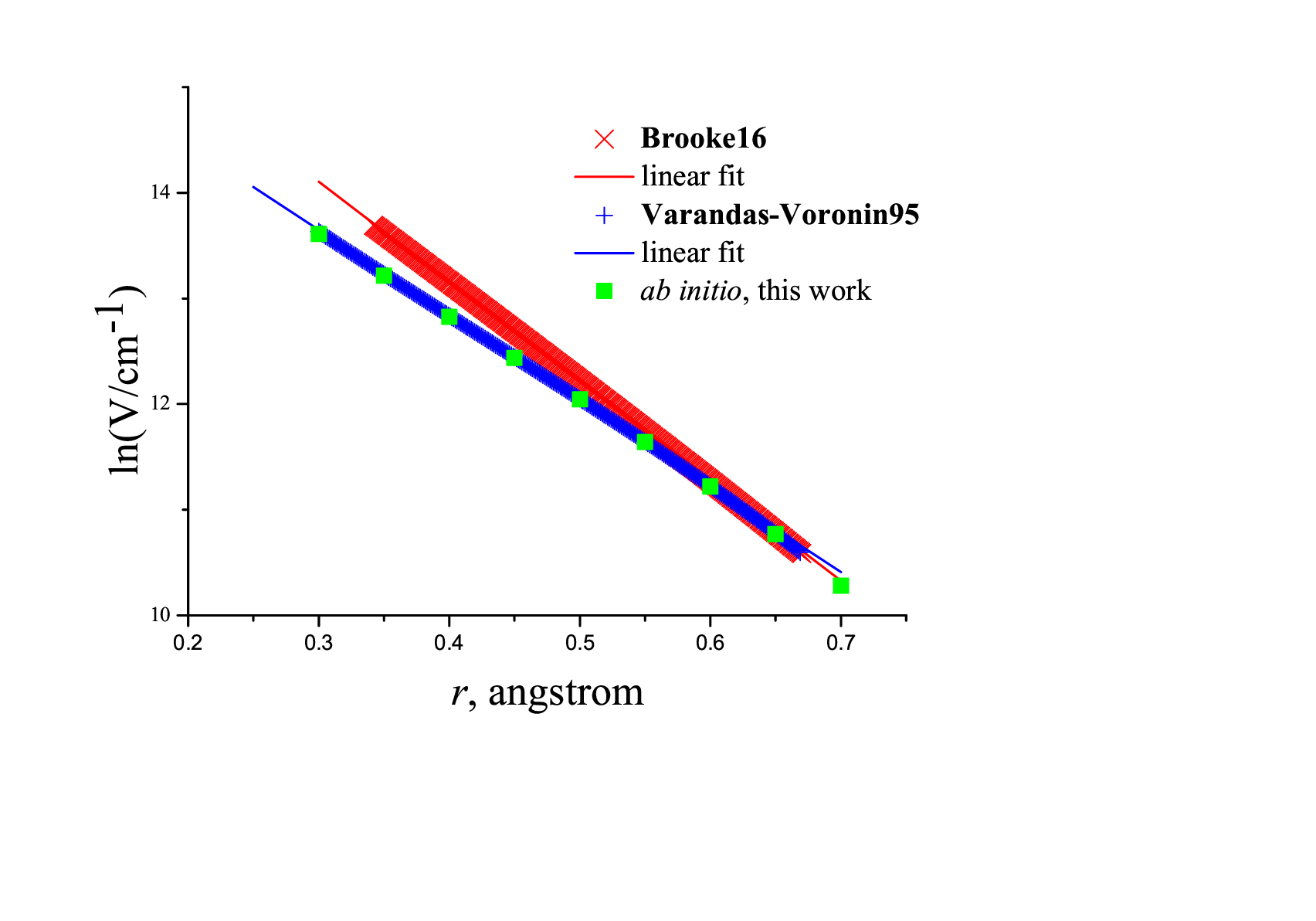}
    \caption{The repulsive branches of two PEFs with linear fits, which give $\beta=4.72\pm0.01$ \AA$^{-1}$ for the \textbf{Brooke16} \cite{Brooke16} PEF and $4.05\pm0.02$ \AA$^{-1}$ for the \textbf{Varandas-Voronin95} \cite{Varandas95} PEF.}
    \label{fig:repulsive}
\end{figure}

Thus, we have got the NIDL slopes in the range of 5.22-5.88, which results in the $\beta$ values in the range of (3.37-3.79) \AA{}$^{-1}$. Taking the largest statistical error of $\delta a=\pm0.48$, we obtain $\delta\beta=(\beta/a)\delta a=\pm0.35$ \AA{}$^{-1}$. This results will be compared with the available data on the repulsive branch of PEF.

The PEF of \textbf{Brooke16} is based on the RKR points and is presented in the tabular form in Supplementary material to Ref. \cite{Brooke16}. One more PEF was developed by Varandas and Voronin in an analytic form \cite{Varandas95}. The \textbf{Varandas-Voronin95} \cite{Varandas95} potential was constructed so as to reproduce the asymptotic united- and separated-atoms limits. Figure \ref{fig:repulsive} shows the repulsive branches of both these potentials. The boudaries of the repulsive branch are determined as follows: it must extend from the left turning point of the $v=2$ level (0.67 \AA) down to a point $r^\star$ where the PEF reaches the value of $D_\textrm{e}\left(v_\textrm{max}/2\right)^2$ where $v_\textrm{max}$ is the desirable upper level \cite{Medvedev12}. Taking $v_\textrm{max}=10$, we obtain $r^\star=0.34$ \AA\ for the \textbf{Brooke16} PEF and 0.3 \AA\ for the \textbf{Varandas-Voronin95} PEF.

The $\beta$ values obtained from the linear fits to both the PEFs are indicated in the figure caption, they are significantly larger that the above estimates. Inserting them into Eq. (\ref{a}), we obtain that the NIDLs associated with these two PEF's repulsive branches must have the slopes of $a=4.19\pm0.01$ and $4.89\pm0.02$, respectively, which are appreciably lower than those obtained from the NIDL plots shown in Tables \ref{tab:eq1}-\ref{tab:eq3_1}. Thus we conclude that \emph{Eq. (\ref{a}) fails} in the present case.

We have already encountered such a situation in molecular hydrogen \cite{Ushakov24}, where it was found that the prefactor in the quasi-classical expression for the TDM is NOT a slow function of the vibrational quantum numbers, as was the case in CO \cite{Medvedev22a} or PN \cite{Ushakov23PN}. It was shown that this is due to the small reduced mass of H$_2$, which entails relatively small actions (on the order of 20 in the $\hbar$ units) as compared to heavier molecules ($\sim1500$). OH also has a small mass, hence, the prefactor also contributes to the NIDL slope. 

In summary, \emph{in the H-containing diatomics, the NIDL slope derived from the calculated and/or observed intensities will always be larger than the one predicted from the steepness of the repulsive branch by Eq. (\ref{a})}.

\section{The \ai\ calculations of the repulsive branch}\label{ai}

In Fig. \ref{fig:repulsive}, it is seen that the repulsive branches of the two PEFs are essentially different. Since the steepness of the repulsive branch affects the overtone intensities \cite{Medvedev12}, it is interesting to learn which one of these PEFs has correct $\beta$. 

To this end, we performed \ai\ calculations of the PEFs of the $X^2\Pi$ and $A^2\Sigma^+$ electronic states by the MRDCI approach. The active space of the MOs is prepared by the CASSCF method with the state averaging technique\footnote{Averaging over two $\Pi$ and one $\Sigma$ states.} for one double occupied core MO and 7 electrons distributed over 9 active MOs. The total amount of configurations in MRDCI are about 6 millions for each of the $X$ and $A$ states. The aug-cc-pV5Z basis set is used at the O and H atoms. All calculations are carried out by GAMESS-US program package \cite{GAMESS}.

The results are shown in Fig. \ref{fig:repulsive}.\footnote{The \ai\ data for the repulsive branch of the $X^2\Pi$ state are given in Supplementary material.} It is seen that the \ai\ points perfectly follow the \textbf{Varandas-Voronin95} PEF. This result is very significant because the analytical results of Varandas and Voronin are confirmed by the first-principles calculation. From the point of view of the NIDL theory, this means that the analytical \textbf{Varandas-Voronin95} PEF is more suited for calculations of the overtone intensities than the point-wise \textbf{Brooke16} one since it has correct repulsive branch.
Therefore, the \textbf{Varandas-Voronin95} potential is a good candidate for calculations of the overtone intensities. It should be noted that the \textbf{Brooke16} PEF is perfectly adjusted to describe the line positions. However, it cannot be excluded that the best description of the overtone intensities will be reached with a different PEF.

\section{Conclusions}\label{conclusions}

With use of the NIDL, we performed the analysis of the intensities calculated by Brooke \emph{et al.} \cite{Brooke16} and found that the intensities of the $\Delta v>6$ transitions should not be trusted, in accord with our preliminary estimates \cite{Medvedev22b}.

The analysis of the OH data revealed one more limitation to the NIDL theory, in addition to the anomalies. Namely, the theory is not applicable to the satellite bands.

The effect of the small reduced mass, previously discovered in H$_2$ \cite{Ushakov24}, causing the NIDL straight-line slope to be larger than the one associated with the repulsive branch of the potential, is demonstrated for OH as well. The same should be true of all the diatomic hydrides, X-H. 

Finally, we performed \ai\ calculations of the OH repulsive branch and compared it with two OH potentials, one (point-wise, spline-interpolated) PEF of Brooke \emph{et al.} \cite{Brooke16} and the other (analytical) PEF due to Varandas and Voronin \cite{Varandas95}. We found that the \ai\ PEF closely follows the \textbf{Varandas-Voronin95} potential in the repulsive region, which is not surprising since the latter was specially constructed to correctly describe both asymptotic limits, in particular the united-atom limit so important in determination of the overtone intensities \cite{Medvedev12}. On the other hand, the \textbf{Brooke16} potential is perfectly suited to describe the line positions. Therefore, we dare to assume that, when selecting data for the spectroscopic databases like HITRAN, HITEMP, \emph{etc.}, different potentials can be used to calculate transition frequencies and transition intensities.

\section*{Acknowledgement}

This work was performed under a ``FRC of PCP and MC RAS" state task, state registration number 124013000760-0.

\bibliography{Mol_Phys_OH}
\bibliographystyle{elsarticle-num}



\bigskip

\end{document}